\documentstyle [12pt,a4p,epsfig,amsmath,multicol]{article}
\newcommand{\PT}{{\rm PT}}
\newcommand{\I}{{\rm I}}
\newcommand{\IP}{{\rm I'}}

\begin{document}
%\hspace*{7cm} {UGVA-DPNC 2003/2-188 February 2003}
%\hspace*{6cm} {UGVA-DPNC 2003/11-188 February 2003}
%\newline
%\hspace*{10cm} {hep-ph/0301231}
%\hspace*{9cm} {hep-ph/0301231}
%\hspace*{8cm} {UGVA-DPNC 1998/07-177 July 1998}
%\newline
% {\it Revised and corrected version of UGVA-DPNC 1997/10-172, CERN OPEN 97-033,
% hep-ph/9801355.  To be published in Modern Physics Letters A.}
\vspace*{0.6cm}

\begin{center} 
{\normalsize\bf Quantum electrodynamics and experiment demonstrate the 
  non-retarded nature of electrodynamical force fields}
\end{center}
\vspace*{0.6cm}
\centerline{\footnotesize J.H.Field}
\baselineskip=13pt
\centerline{\footnotesize\it D\'{e}partement de Physique Nucl\'{e}aire et 
 Corpusculaire, Universit\'{e} de Gen\`{e}ve}
\baselineskip=12pt
\centerline{\footnotesize\it 24, quai Ernest-Ansermet CH-1211Gen\`{e}ve 4. }
\centerline{\footnotesize E-mail: john.field@cern.ch}
\baselineskip=13pt
 
\vspace*{0.9cm}
\abstract{In quantum electrodynamics, the quantitatively most successful
  theory in the history of science, intercharge forces obeying
 the inverse square law are due to the
   exchange of space-like virtual photons. The fundamental quantum process 
   underlying applications as diverse as the gyromagnetic ratio of the
   electron and electrical machinery is then M\o ller scattering $ee \rightarrow ee$.
   Analysis of the quantum amplitude for this process shows that the corresponding
    intercharge force acts instantaneously. This prediction has been 
    verified in a recent experiment. }

\vspace*{0.9cm}
\normalsize\baselineskip=15pt
\setcounter{footnote}{0}
\renewcommand{\thefootnote}{\alph{footnote}}
%\newline
%PACS 03.65.Bz, 14.60.Pq, 14.60.Lm, 13.20.Cz 
%\newline
%{\it Keywords ;} Quantum Mechanics,
%Neutrino Oscillations.
%\newline
%\newline 
%{\it Keywords ;} Standard Electroweak Model, LEP and SLD data, Z-decays,
% Anomalous right-handed b quark coupling.
%\newline
%\vspace*{0.4cm}

 \par One of the most remarkable developments of physical science during the 20th
   Century was the advent of quantum electrodynamics (QED) ~\cite{KinQED} born of
   the fusion of quantum mechanics and special relativity theory, QED has two important
    aspects; the first is the remarkable quantitative success of the theory --an essentially
    perfect description of nature within its domain of applicabilty. The second is its role
    as a `model theory' from which the standard model of particle physics was later developed
    by attempting to describe the weak and strong interactions by quantum field theories
    in a similar way as the electromagnetic interaction is described by QED.
    A recent illustration of the former aspect of QED is the remeasurement at
     Harvard University of the gyromagnetic ratio of the electron~\cite{OHUG}:
     \[ \frac{g_e^{exp}}{2} = 1.00115965218085(76)  \]
     The measurement uncertainty in the last two figures is indicated in parentheses.
      The accuracy of the measurement is 7.6 in $10^{13}$, a six-fold improvement
      on the previous best measurement performed at the University of Washington in 1987~\cite{UWGM2}
       for which Hans Dehmelt was awarded a part of the 1989 Nobel Prize in Physics. The electron 
     gyromagnetic ratio is the most precisely measured physical quantity, to date, in the history
      of science. The QED prediction for $g_e$ is~\cite{GHKNO}:
      \[ \frac{g_e^{thy}}{2} = 1.00115965218875(766)  \]
       Where the measured value of the fine structure constant: $\alpha = 1/137.036...$ derived 
     from spectroscopy of the Rubidium atom~\cite{Rubid}, is used in the prediction. Theory and experiment
      for $g_e$ agree at about one part in $10^{11}$~\cite{GHKNO}:
      \[  \frac{g_e^{exp}}{2} - \frac{g_e^{thy}}{2} = -7.9(7.7) \times 10^{-12} \]
       Assuming the correctness of the QED prediction for $g_e$, the Harvard measurement determines
       the value of $\alpha$ with ten times better accuracy than given by  Rubidium spectroscopy, which
        yields the second most accurate experimental measurement of this constant.
       \par The calculational problems which must be surmounted in order to obtain the QED
            prediction for $g_e$ at the accuracy required by the latest experimental measurement, and so
             obtain a quantitatively meaningful comparison of theory and experiment, are formidable.
            However, as pointed out by Feynman, the basic physical concepts on which the calculations
            are based are extremely simple~\cite{FeynQED}. Three fundamental quantum mechanical 
            amplitudes suffice to calculate the amplitude of any QED space-time process, no matter
            how complicated, involving
            only structureless charged particles and photons. These are the amplitudes for the processes:\newline
                 \par A photon goes from place to place \newline
                  \par A charged particle goes from place to place \newline
                 \par A charge particle emits or absorbs a photon \newline
              \par Combination of these three amplitudes enables the quantum mechanical
             probability amplitude for any space-time process in QED --one Feynman
             diagram per amplitude--to be written down. Depending on the accuracy with
        which it is desired to obtain the prediction for the value of some measured observable,
         quantum mechanical superposition must be used to add the amplitudes corresponding to
         one or more Feynman diagrams. The accuracy of the latest Harvard
         measurement of  $g_e$ requires the evaluation of the amplitudes of 891 distinct
          Feynman diagrams containing up to five virtual photon lines.
           \par The lowest order diagram contributing to the `anomaly' $a_e\equiv g_e/2-1$
         is shown in Fig.1a. The physical observable corresponding to this diagram
         is the angular frequency with which the spin vector of the electron
         rotates relative to the direction of its momentum vector: $\omega_a = a_e B/m_e$. 
         The diagram of Fig.1a gives the prediction $a_e = \alpha/\pi$~\cite{Schwinger}.
         The left diagram in Fig.1a shows the conventional way of drawing the diagram,
         where the vertical virtual photon line is identified with the (classical) uniform
         magnetic field $B$ produced, in the Harvard experiment, by the solenoidal magnet of a
          Penning Trap (PT). In this experiment it is the
           gyromagnetic ratio of a single electron, $e_{\PT}$, initially in the lowest
          lying cyclotron level of the PT, which is measured. In the QED description
          the classical magnetic field, $B$, is a manifestation of the exchange of
         virtual photons between the conduction electrons, $e_{\I}$, of the solenoidal
         magnet and  $e_{\PT}$. This is shown in the diagram on the right in Fig.1a.
           The field $B$ is the result, in QED, of the superposition of the amplitudes
           corresponding to each conduction electron in the magnet. Under the influence
        of the transverse Lorentz force generated by $B$,  $e_{\PT}$ undergoes
        periodic cyclotron motion in the PT with angular frequency  $\omega_c =  e B/m_e$.
        The corresponding diagram in the QED description is shown in Fig.1b.
         The Lorentz force is produced by the exchange of virtual photons between
         $e_{\PT}$ and all the conduction electrons in the magnet. Using the technique
        of `Quantum-Jump Spectroscopy'~\cite{PT}, the Harvard experiment measures separately
         $\omega_a$ and $\omega_c$ and derives $a_e$, and hence $g_e$, from the relation
           $a_e = \omega_a/\omega_c$. 
\begin{figure}[htbp]
\begin{center}
\hspace*{-0.5cm}\mbox{
\epsfysize15.0cm\epsffile{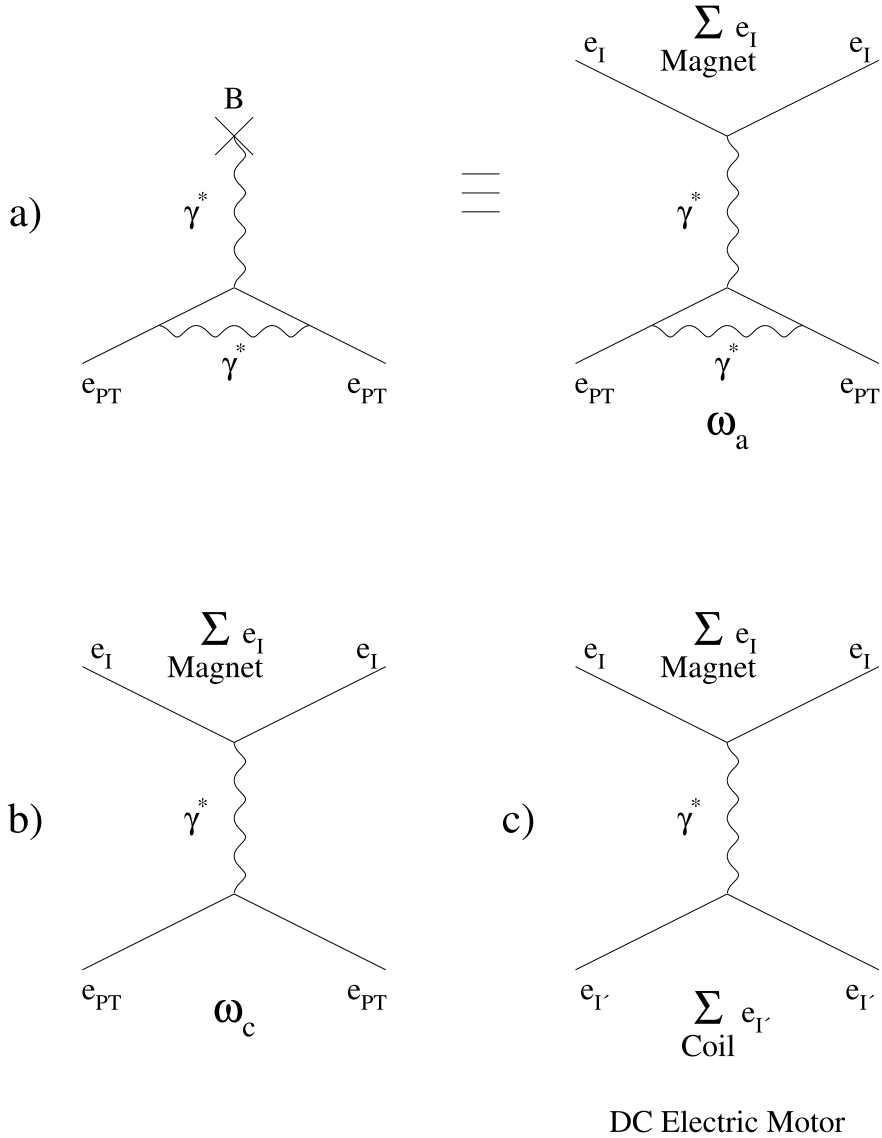}}   
%\mbox{\epsfig{file=qcdf1.eps,height=12cm}}
%\mbox{\epsfig{file=test_new.eps,height=6cm}}
\caption{{\sl Lowest order Feynman diagrams contributing to: a) $a_e =g_e/2-1$ or the 
      spin precession frequency $\omega_a$, b) the cyclotron frequency  $\omega_c$ in the Penning
      Trap of Ref.~\cite{OHUG}, c) the operation of a direct current electric motor.
       See text for discussion}}
\label{fig-fig1}
\end{center}
 \end{figure}
          \par The force on  $e_{\PT}$  in Fig.1b, produced by virtual photon exchange,
             that causes the cyclotron motion of the trapped electron has the same
             origin, in QED, as the force which causes the coil of a DC electric
                motor to rotate (Fig.1c). The electron  $e_{\PT}$  is simply 
            replaced by a conduction electron $e_{\IP}$ in the rotating coil of the motor.
          Each pair of conduction electrons, one in the magnet, one in the coil,
           gives a contribution to the turning force of the motor. 
          \par The fundamental QED process underlying cyclotron motion of  $e_{\PT}$
            in the PT of the Harvard experiment, an electric motor, and indeed all
             inter-electron forces obeying an inverse square law in electrodynamics, including those operating
             in the experiment to be described below in the present paper, is then 
              M\o ller scattering: $e e \rightarrow e e$ as in Fig.1b and 1c.
             The lowest order Feynman diagram for the process shown involves the
             exchange of a single `space-like' virtual photon. The meaning of this
             appellation is explained below.
         \par In momentum space, the invariant amplitude for M\o ller scattering
            $e_Ae_B \rightarrow e_Ae_B$ is~\cite{HM} 
     \begin{equation} 
       T_{fi} =  -i \int \frac{{\cal J}^A(x_A) \cdot {\cal J}^B(x_A)}{q^2} d^4x_A
    \end{equation}
         The 4-vector current ${\cal J}^A$ is defined in terms of plane-wave solutions,
      $u_i$,~$u_f$ of the Dirac 
     equation for the incoming ($i$) and outgoing ($f$) electron $e_A$ as
   \begin{equation}
    {\cal J}^A_{\mu} \equiv - e \overline{u}_f^A \gamma_{\mu} u_i^A \exp[i(p_f^A-p_i^A)\cdot x_A]
     \end{equation} 
      The incoming electron of 4-vector momentum $p_i^A$ emits the virtual photon at the space-time
        point $x_A$ and scatters with  4-vector momentum $p_f^A$. The factor $1/q^2$ in (1) is the
        amplitude, in momentum space, for the virtual photon to propagate between points
       on the trajectories of  $e_A$ and $e_B$, $q$ being the 4-vector momentum of the
         virtual photon. The integral over $ d^4x_A$ in (1) in conjunction with the
            exponential functions in the currents (equivalent to Dirac-$\delta$ functions)
            ensures energy-momentum conservation in the scattering process.
           In the overall center of mass (CM) frame, energy-momentum conservation
        requires that the energy, but, in general, not the momentum, of the virtual 
           photon vanishes. This implies that in the CM frame:
   \begin{equation}
        q^2 = q_0^2-|\vec{q}|^2 \rightarrow -|\vec{q}|^2
   \end{equation}
      The virtual photon is then `space-like' because its  4-momentum squared is negative.
       If an object has positive 4-vector momentum squared, its 4-vector momentum is `time-like' and a
      positive mass
        may be assigned to it. All ponderable physical objects in the real world
       have such time-like 4-vector momenta. As pointed out by Einstein in 1905~\cite{Ein1}, the
        speed of an object with a time-like  4-vector momentum is less than that of light. 
        However, as just shown, the virtual photons exchanged in  M\o ller scattering are not time-like.
       Using (3), (1) may be written, in the CM frame, as:
     \begin{equation} 
       T_{fi} = i \int \frac{{\cal J}^A(x_A) \cdot {\cal J}^B(x_A)}{|\vec{q}|^2} d^4x_A
       \end{equation}
        The Fourier transform:
       \begin{equation} 
        \frac{1}{|\vec{q}|^2} = \frac{1}{4 \pi} \int \frac{d^3 x e^{i\vec{q} \cdot \vec{x}}}{|\vec{x}|}
          \end{equation}
 enables the invariant amplitude to be written as a space-time integral~\cite{JHFRCED}:

   \begin{equation} 
    T_{fi} = \frac{i}{4 \pi} \int dt_A \int d^3 x_A  \int d^3 x_B \frac{{\cal J}^A(\vec{x}_A,t_A) \cdot
     {\cal J}^B(\vec{x}_B,t_A)}{|\vec{x}_B-\vec{x}_A|} 
           \end{equation} 
     It can be seen that the momentum-space photon propagator $1/|\vec{q}|^2$, the value of
      which is fixed by the electron scattering angle, corresponds, in space-time, to the exchange
      of an infinite number of virtual photons travelling between all spatial positions
       $\vec{x}_A$ and $\vec{x}_B$ on the trajectories of the electrons $e_A$ and $e_B$. 
         Eqn(6) shows that {\it each virtual photon is both emitted and absorbed at the same instant,
        so that the corresponding force is  transmitted instantaneously}. The same conclusion
        follows from considerations of relativistic kinematics. The magnitude of the velocity
        of an object with momentum $p$ and energy $E$ is $v = p c^2/E$. In the CM frame of 
        M\o ller scattering, the virtual photon has $p > 0$ for any non-zero scattering angle,
        but always $E = 0$. Therefore $v$ is infinite and the corresponding intercharge interaction
        instantaneous. 
        \par It has been previously noticed that the `Coulomb interaction' associated with 
              the temporal components of the currents  ${\cal J}^A$  and  ${\cal J}^B$
            (exchange of `longitudinal' virtual photons) is instantaneous, but it is usually
             stated that the interaction mediated by the spatial components of the currents
            (exchange of `transverse' virtual photons) is transmitted at the speed of light
            ~\cite{FeynRI}. The arguments presented above show instead that the whole
           intercharge interaction is instantaneous in the CM frame. 
        \par QED provides a simple explanation for Coulomb's inverse square law of force,
         which, in this theory, is mediated by the exchange of virtual photons. Geometrical
           considerations and conservation of particle number implies that, if a force
          is proportional to the number of some exchanged particles, it must decrease as the inverse
          square of the distance from the source. An example is the force of radiation 
           pressure due to real photons emitted from the Sun. Quantum mechanics, in
          general, modifies this prediction. For a virtual particle of pole mass $m$
            propagating over a large space-like interval $\Delta x > \Delta t$ the 
          amplitude for displacements  $\Delta x$, $\Delta t$ (the space-time propagator)
          takes the form: $\exp[-m\sqrt{(\Delta x)^2-(\Delta t)^2}]$~\cite{FeynFP}.
          Thus the range of the force is exponentially damped if $m$ is non zero.
          Since, however, a photon has vanishing pole mass, no damping 
          occurs for the case of space-like virtual photons, so the inverse square law
          is expected to hold. The same argument applies to the propagator of any
           other massless, particle, independantly of its interactions; e.g. exchange
          of a space-like virtual massless graviton could explain the inverse square law of the
          gravitational force. 
         \par The above predictions of QED: an instantaneous intercharge interaction and
           the Coulomb inverse square law in electrostatics, have been used, in conjunction
          with relativistic invariance and Hamilton's Principle, to formulate a classical
          theory of interchange forces without {\it a priori} introduction of any
           classical field concept~\cite{JHFRCED}. The dynamical equations of this theory,
            Relativistic Classical Electrodynamics (RCED) are derived by  Hamilton's Principle
           from the Lorentz-invariant Lagrangian:
          \begin{equation}
        L(x_1,u_1;x_2,u_2) = -\frac{m_1 u_1^2}{2} -\frac{m_2 u_2^2}{2}
        - \frac{j_1 \cdot j_2}{c^2  \sqrt{-(x_1-x_2)^2}}
         \end{equation} 
          that describes the electromagnetic interaction between two charged objects O1,O2
          with space-time, velocity and current 4-vectors $x$, $u$ and $j$ respectively, and
          Newtonian mass $m$.
           The fields and potentials of Classical Electromagnetism (CEM), are all derived,
           as well as relativistic corrections to them, by mathematical substitition , in the
           dynamical equations obtained by inserting the Lagrangian (7) into the Lagrange
           equations that follow from Hamilton's Principle. Also predicted are the
           Faraday-Lenz law, the Biot and Savart law and the Lorentz force law.  The complete
           relativistic description of the effect of inverse square intercharge forces between O1 and O2
           is provided by the `fieldless' first-order differential equations:
                \begin{eqnarray} 
      \frac{d\vec{p_1}}{dt} &  = & \frac{q_1}{c}\left[\frac{ j_2^0\vec{r} +  \vec{\beta}_1 \times
     (\vec{j_2} \times \vec{r})}{r^3} -\frac{1}{c r}\frac{d \vec{j_2}}{d t}-\vec{j_2}
     \frac{(\vec{r} \cdot \vec{\beta}_2)}{r^3} 
      \right] \\
       \frac{d\vec{p_2}}{dt}  &  = & -\frac{q_2}{c}\left[\frac{ j_1^0\vec{r} +\vec{\beta}_2 \times 
%  (\vec{j_1} \times \vec{r})}{r^3}+\frac{1}{c r}\frac{d \vec{j_2}}{d t} -\vec{j_1}
  (\vec{j_1} \times \vec{r})}{r^3}+\frac{1}{c r}\frac{d \vec{j_1
}}{d t} -\vec{j_1}
     \frac{(\vec{r} \cdot \vec{\beta}_1)}{r^3}
    \right] 
     \end{eqnarray} 
         where $r \equiv |\vec{x}_1-\vec{x}_2|$, $\beta \equiv v/c$. 
       On neglecting relativistic corrections of O($\beta^2$) and higher, these equations
       are equivalent to Coulomb's law, the Biot and Savart Law  and the Lorentz force law
       of CEM.  
  \par Not described by (8) and (9) are the effects of {\it real} photons propagated at the speed
   of light, $c$. The corresponding {\it radiative} electric and magnetic fields, to be 
     contrasted with the {\it force fields} implicit in (8) and (9), are produced by
    accelerated charges and have an $r^{-1}$ dependence, instead of the 
      $r^{-2}$ dependence of the force fields. On solving the coupled differential
      equations (8) and (9), for the particular case of circular Keplerian orbits of
      two equal and opposite charges~\cite{JHFRCED}, it is found that the $r^{-1}$
      dependent terms in (8) and (9), although also containing acceleration factors,
      do not describe radiation of real photons, but rather the modification of the masses
      of the objects due to their mutual electromagnetic interaction.
   \par In any experiment where source charges are accelerated and the effect on test charges
       is observed, contributions of two distinct types are therefore to be expected:
       \begin{itemize}
          \item[(i)] The effect of instantaneous {\it force fields}, mediated in QED by the
                 exchange of space-like virtual photons, with $r^{-2}$ dependence.
          \item[(ii)] The effect of {\it radiation fields}, mediated by propagation
             at the speed of light of real (on shell) photons, with
          with $r^{-1}$ dependence.
       \end{itemize}
      Because of the rapid fall-off with distance of the force fields, special
        experimental arrangements are necessary for their detection. 
       Remarkably, it is only this year, more than a century after Hertz' experiment~\cite{Hertz}
      in which `electromagnetic waves' were discovered, that the results of a new one sensitive to the 
      temporal properties (instantaneous or retarded) of the force fields has been
       published~\cite{KMSRIC}. This experiment is now briefly described.
        \par In essence, the experiment is a repetition of the Hertz experiment using modern
           electronics to detect and visualise the signals, and probing small separations
        of the emitting and receiving antennas in order to be sensitive to the force fields
        with  $r^{-2}$ dependence (called by the authors of Ref.~\cite{KMSRIC},
        `bound fields'). The emitting (EA) and receiving (RA) antennae are essentially
         one-turn circular
         coils of radius 5cm and depths 5cm(EA) and 10cm(RA) consisting of 1mm thick copper sheet,
         and placed in the same horizontal plane, with centers separated by the distance $R$. The EA
         is activated by discharging a capacitor $C$, in series with the EA of inductance $L$, with the
         aid of a spark gap, to generate a pulsed harmonically varying current with
         angular frequency $\omega = 1/\sqrt{LC} = 7.4 \times 10^8$ rad/sec. The associated
         time-varying magnetic field induces a current in the RA which is displayed as a temporal
         signal on a 500Mhz digital oscilloscope. 
         \par Neglecting relativistic corrections, the magnetic field produced by a small single-turn
          coil of surface area $\Delta S$ containing a current $I$ at a large distance $R$ from
          the coil in the plane of the latter is (see the Appendix  of Ref.~\cite{KMSRIC}):
           \begin{equation}
          \vec{B} = \vec{B}_v^{force}+\vec{B}_c^{rad} =
                \frac{\Delta S}{4 \pi \epsilon_0 c^2} \left\{
                     \frac{[\I]_v}{R^3} +\frac{c}{v}\frac{[\dot{\I}]_v}{c R^2} +\frac{[\ddot{\I}]_c}{c^2 R}
                 \right\} \hat{k}
           \end{equation}
          where the dot denotes a time derivative and the unit vector $\hat{k}$ is perpendicular to the
          plane of the coil.  The terms with $R^{-3}$ and  $R^{-2}$ dependence describe the effects of the
          force field, and are derived from the Biot and Savart law, that with  $R^{-1}$ dependence is a
          radiation field. The square 
         brackets indicate retardation of the enclosed quantity; that is, it is evaluated at the
         time $t-R/v$ for the force field and  $t-R/c$ for the radiation field. In CEM it
         is usually assumed that $v = c$, whereas QED predicts that
         $v = \infty$. In the experiment, $v$ is assigned an arbitary value in (10) and measured
          by comparing the prediction of this formula with the experimental data..
         The Faraday-Lenz law predicts that the time varying current signal, $\epsilon_v(t)$,
        in a parallel test coil, also
         of surface area $\Delta S$, placed in the field $\vec{B}$ is:
           \begin{equation}
         \epsilon_v(t) =
                \frac{(\Delta S)^2}{4 \pi \epsilon_0 c^2} \left\{
                \frac{[\dot{\I}]_v}{R^3} +\frac{c}{v}\frac{[\ddot{\I}]_v}{c R^2} +\frac{[\dddot{\I}]_c}{c^2 R}
                 \right\} 
           \end{equation}
          The harmonic variation of the magnetic field produced by EA implies that (11) may be written as:
           \begin{equation}         
    \epsilon_v(t) = \epsilon_0\left[-\frac{\sin\omega(t-R/v)}{R^3} + \frac{\omega \sin[\omega(t-R/v) -\pi/2]}
                 {vR^2} + \frac{\omega^2 \sin\omega(t-R/c)}{c^2 R} \right]
       \end{equation}
     It can be seen that the phases of the force field contributions in the first two terms are different to that
       of the radiation field contribution in the third term on the right side of (12). Since the phase
       difference between the force field and radiation field contributions depends on both $R$ and $v$,
      the value of $v$ can be determined by measuring the $R$ dependence of the signal function $\epsilon_v(t)$.
       In practice this is done by measuring the $R$ dependence of the time difference, $\Delta t$,
         between the first zero crossing points of the total signal  $\epsilon_v(t)$
        and a reference signal, $\epsilon^{ref}(t)$, provided by measuring  $\epsilon_v(t)$ at large distances
        where only
        the radiative contribution remains:
           \begin{equation}         
   \epsilon^{ref}(t) \equiv  \epsilon^{rad}(t) =  \epsilon_0 \frac{\omega^2 \sin\omega(t-R/v)}{c^2 R}
       \end{equation}
     and extrapolating it back to the short distance region. 
     In order to reproduce correctly the experimental conditions, the elementary formula (11) was
    numerically integrated over the surfaces and depths of the EA and the AR.
\begin{figure}[htbp]
\begin{center}
\hspace*{-0.5cm}\mbox{
\epsfysize15.0cm\epsffile{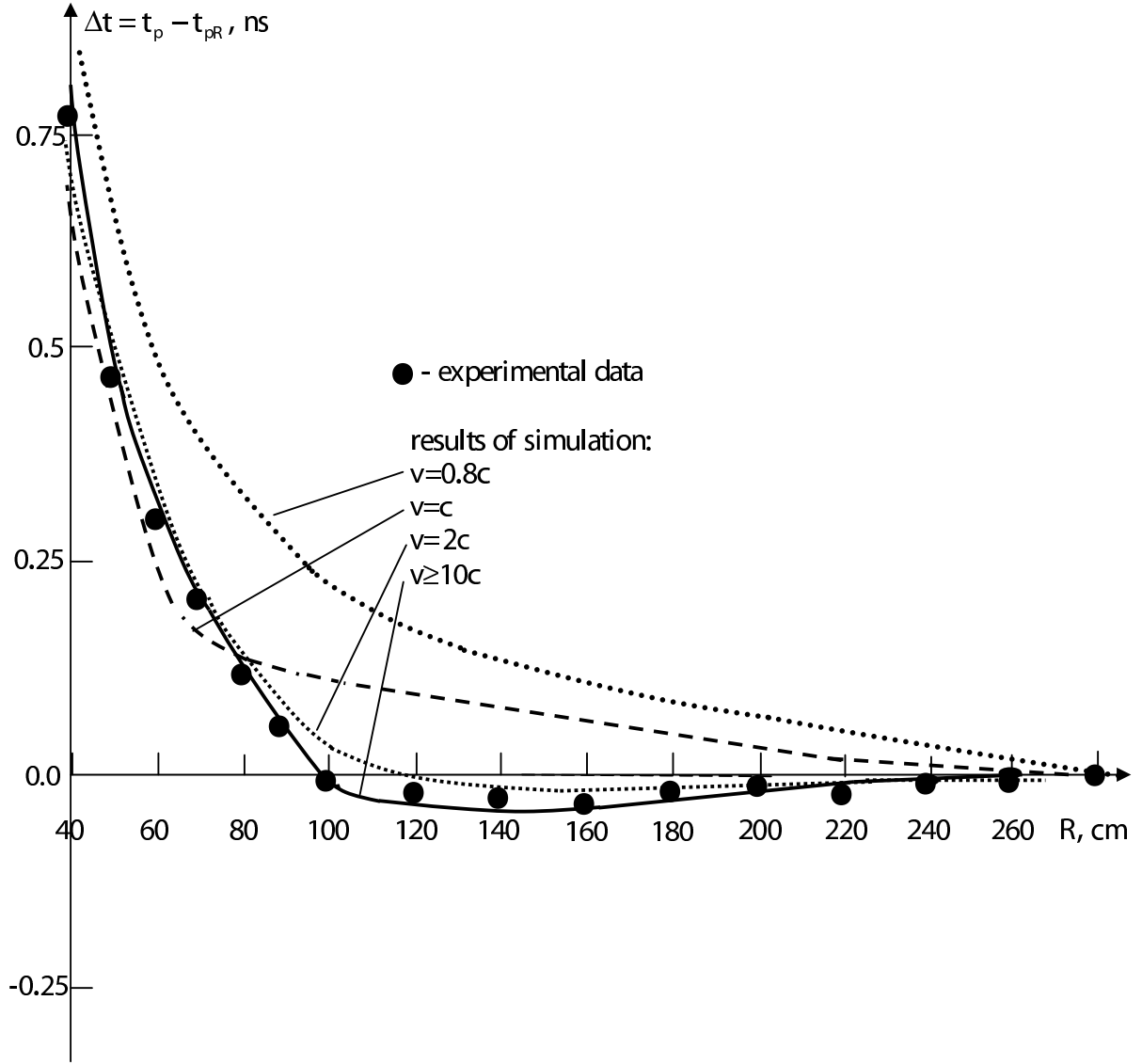}}   
%\mbox{\epsfig{file=qcdf1.eps,height=12cm}}
%\mbox{\epsfig{file=test_new.eps,height=6cm}}
\caption{{\sl Measured values of the difference, $\Delta t$, between the
    first zero crossing times of the signals $\epsilon_v(t)$ and $\epsilon^{ref}(t)$,
     in comparison with simulations for different values of $v$ from Ref.\cite{KMSRIC}. The measurement error
    on each experimental datum is $\simeq 0.02$ns. The prediction $v = c$ of conventional CEM
    is completely excluded whereas the data are consistent with the QED prediction
     $v = \infty$ which is indistingushable from the $v \ge 10c$ curve shown.}}
\label{fig-fig2}
\end{center}
 \end{figure}
    \par The measured results for $\Delta t$ as a function of $R$, in comparison with predictions
      for different values of $v$, are shown in Fig.2. It can be seen that the case of conventionally
      retarded force fields, as in CEM, is completely excluded by the measurements, but that good agreement
      is found for the case $v\ge 10c$, which is essentially the same as the QED prediction $v = \infty$.
      Yet again a prediction of QED is in perfect agreement with experiment!
        \par The physical paradigm of `causality' --that no physical influence can propagate faster
        than the speed of light -- universally accepted in physics in the second half of the 19th
        Century and the whole of the 20th-- is thus in contradiction both with the prediction of QED
        and the experimental results shown in Fig.2 --the force fields generated by the EA arrive at
        the RA before the radiation fields, in QED the virtual photons arrive before the 
        real ones. Einstein's argument in the 1905 special relativity paper~\cite{Ein1} that 
       `causality' follows from special relativity is correct, only insofar as `information' is transmitted
       by objects with time-like 4-momentum vectors, However the virtual photons that manifest
       as the force fields of CEM or RCED have space-like, not time-like 4-vectors. Any particle
       with a space-like 4-momentum vector is tachyonic in nature --the speed of light is
       a lower, not an upper, limit on its speed.
       \par It is interesting to notice that Hertz' paper~\cite{Hertz} reporting the discovery
        of electromagnetic waves propagating with a velocity `akin to that of light' also showed data,
        taken close to the source, that was consistent with an infinite propagation speed for the
         associated signal~\cite{SR,Buchwald}. That no comment was made on this data in the conclusions
         of Ref.~\cite{Hertz} is the more surprising, given that two types of electric force, one
         instantaneous, the other propagating at the speed of light were proposed in the 
         electromagnetic theory of Hertz' mentor, Helmholtz~\cite{SR}.  
       \par The author thanks Professor A.L.Kholmetskii for bringing Ref.\cite{KMSRIC} to his attention,
        and for providing Fig.2.

%\pagebreak

\end{document}